\documentclass[english,nofootinbib,aps,reprint,superscriptaddress,twocolumn]{revtex4-2}
\usepackage[T1]{fontenc}
\usepackage[latin9]{inputenc}
\usepackage{xcolor}
\usepackage{array}
\usepackage{booktabs}
\usepackage{amsmath}
\usepackage{amssymb}
\usepackage{graphicx}
\usepackage{esint}
\PassOptionsToPackage{normalem}{ulem}
\usepackage{ulem}

\makeatletter

\providecommand{\tabularnewline}{\\}
\providecolor{lyxadded}{rgb}{0,0,1}
\providecolor{lyxdeleted}{rgb}{1,0,0}
\DeclareRobustCommand{\lyxsout}[1]{\ifx\\#1\else\sout{#1}\fi}

\usepackage[english]{babel}
\usepackage{graphicx}% Include figure files
\usepackage{dcolumn}% Align table columns on decimal point
\usepackage{bm}% bold math
\usepackage{xcolor}
\def\urlprefix{}
\def\url#1{}
\addto\captionsenglish{%
}

\usepackage{babel}
\usepackage[none]{hyphenat}

\providecommand*{\diff}%
{\@ifnextchar^{\DIfF}{\DIfF^{}}}
\def\DIfF^#1{%
	\mathop{\mathrm{\mathstrut d}}%
	\nolimits^{#1}\gobblespace}
\def\gobblespace{%
	\futurelet\diffarg\opspace}
\def\opspace{%
	\let\DiffSpace\!%
	\ifx\diffarg(%
	\let\DiffSpace\relax
	\else
	\ifx\diffarg[%
	\let\DiffSpace\relax
	\else
	\ifx\diffarg\{%
	\let\DiffSpace\relax
	\fi\fi\fi\DiffSpace}

\makeatother

\begin{document}
\title{Machine-Learning Recognition of Dzyaloshinskii-Moriya Interaction
from Magnetometry}
\author{Bradley J. Fugetta}
\affiliation{Department of Physics, Georgetown University, Washington, DC 20057,
USA}
\author{Zhijie Chen}
\affiliation{Department of Physics, Georgetown University, Washington, DC 20057,
USA}
\author{Dhritiman Bhattacharya}
\affiliation{Department of Physics, Georgetown University, Washington, DC 20057,
USA}
\author{Kun Yue}
\affiliation{Nvidia Corp., Santa Clara, CA 95051, USA.}
\author{Kai Liu}
\affiliation{Department of Physics, Georgetown University, Washington, DC 20057,
USA}
\author{Amy Y. Liu}
\affiliation{Department of Physics, Georgetown University, Washington, DC 20057,
USA}
\author{Gen Yin}
\thanks{Gen.Yin@georgetown.edu}
\affiliation{Department of Physics, Georgetown University, Washington, DC 20057,
USA}
\begin{abstract}
The Dzyaloshinskii-Moriya interaction (DMI), which is the antisymmetric
part of the exchange interaction between neighboring local spins,
winds the spin manifold and can stabilize non-trivial topological
spin textures. Since topology is a robust information carrier, characterization
techniques that can extract the DMI magnitude are important for the
discovery and optimization of spintronic materials. Existing experimental
techniques for quantitative determination of DMI, such as high-resolution
magnetic imaging of spin textures and measurement of magnon or transport
properties, are time consuming and require specialized instrumentation.
Here we show that a convolutional neural network can extract the DMI
magnitude from minor hysteresis loops, or magnetic \textquotedblleft fingerprints,\textquotedblright{}
of a material. These hysteresis loops are readily available by conventional
magnetometry measurements. This provides a convenient tool to investigate
topological spin textures for next-generation information processing.
\end{abstract}
\maketitle
The Dzyaloshinskii-Moriya interaction (DMI) is an antisymmetric exchange
coupling between neighboring spins, and it is non-zero only in materials
with broken central symmetry \citep{dzyaloshinsky_thermodynamic_1958,moriya_anisotropic_1960}.
The symmetry can be broken intrinsically by the crystal structure,
as in the case of B20 compounds \citep{muhlbauer_skyrmion_2009,neubauer_topological_2009,yu_near_2010,pappas_chiral_2009,yu_real-space_2010,huang_extended_2012},
or it can be broken extrinsically by designing magnetic multilayer
heterostructures \citep{crepieux_dzyaloshinskymoriya_1998,romming_writing_2013,fert_skyrmions_2013,heinze_spontaneous_2011,ferriani_atomic-scale_2008,jiang_blowing_2015}.
Due to its antisymmetric nature, the DMI favors perpendicular configurations
between neighboring spins, winding the spin manifold. As a result,
it plays an essential role for topological spin textures, including
magnetic skyrmions \citep{fert_skyrmions_2013}, vortices\citep{siracusano_magnetic_2016},
bimerons\citep{li_bimeron_2020}, hedgehogs\citep{fujishiro_topological_2019,tanigaki_real-space_2015,okumura_magnetic_2020,kent_creation_2021},
chiral domain walls\citep{chen_novel_2013,ryu_chiral_2013}, and hopfions\citep{tai_static_2018}.
These textures are either directly stabilized by the DMI, or their
behaviors are strongly impacted by the DMI magnitude. Such topological
spin textures have promising potential as information carriers in
next-generation spintronic devices for low-power and high-speed applications\citep{fert_skyrmions_2013,liu_three-dimensional_2020,zou_topological_2020}.
Even in systems with uniform spins, the DMI can induce phenomena such
as non-reciprocal and topological magnon spectra\citep{melcher_linear_1973,udvardi_chiral_2009,costa_spin-orbit_2010},
which are useful for radio-frequency devices. Quantitative understanding
and control of the DMI magnitude in spintronic systems is therefore
important for both fundamental and application purposes.

Significant efforts have been devoted to the quantitative determination
of the DMI magnitude, especially for thin-film multilayers, which
are essential for device applications\citep{kuepferling_measuring_2020}.
In fact, for a thin-film multilayer, the magnitude of the extrinsic
DMI can be continuously modulated by carefully controlling the vertical
profile of the heterostructure\citep{yu_room-temperature_2016,ma_interfacial_2016}.
In addition, tunable control of interfacial DMI has recently been
demonstrated via chemisorption or ionic gating, even after the materials
systems have been synthesized \citep{chen_large_2020,chen_observation_2021,diez_nonvolatile_2019}.
Since details such as chemical bonding and atomic alignment at interfaces
are important, theoretical prediction of extrinsic DMIs from first
principles is very challenging. Experimentally, techniques such as
domain-wall imaging\citep{bode_chiral_2007,ferriani_atomic-scale_2008,heide_dzyaloshinskii-moriya_2008,torrejon_interface_2014},
Brillouin light scattering\citep{zakeri_asymmetric_2010,ma_dzyaloshinskii-moriya_2017,soucaille_probing_2016,ma_interfacial_2016}
and loop-shift measurements using spin-orbit torque setups\citep{torrejon_interface_2014,fukami_magnetization_2016,pai_determination_2016,ding_interfacial_2019,han_room-temperature_2017}
can be used to obtain the DMI magnitude. However, these experiments
are non-trivial and sometimes require trial-and-error iterations between
experiments and modeling.

Recently, predictive machine-learning models have been demonstrated
to successfully extract DMI magnitudes from ground-state spin textures
with good accuracy\citep{kawaguchi_determination_2021,singh_application_2019,kwon_magnetic_2020,wang_machine_2020}.
These approaches typically use phenomenological Hamiltonians considering
leading-order terms, such as the symmetric Heisenberg exchange, the
uniaxial anisotropy, the long-range dipolar interaction, the Zeeman
coupling, and the DMI. To generate the training data, either dynamical
or Monte Carlo simulations are used to find the ground state of a
Hamiltonian given a set of parameters. A neural network can then be
trained to correlate the ground-state spin texture (inputs) and the
Hamiltonian parameters (outputs) including the DMI. The success of
this practice suggests that the ground-state spin texture indeed contains
information about the intricate competition between the DMI and other
terms in the Hamiltonian. This is intuitive since the DMI is the only
term in the Hamiltonian that favors spin winding with a uniform chirality.
However, experimental imaging of spin textures requires high spatial
resolution as well as the ability to resolve the magnetization vectors.
High-quality, exposed pristine surfaces are therefore often necessary.
\begin{figure*}
\begin{centering}
\includegraphics[width=0.85\textwidth]{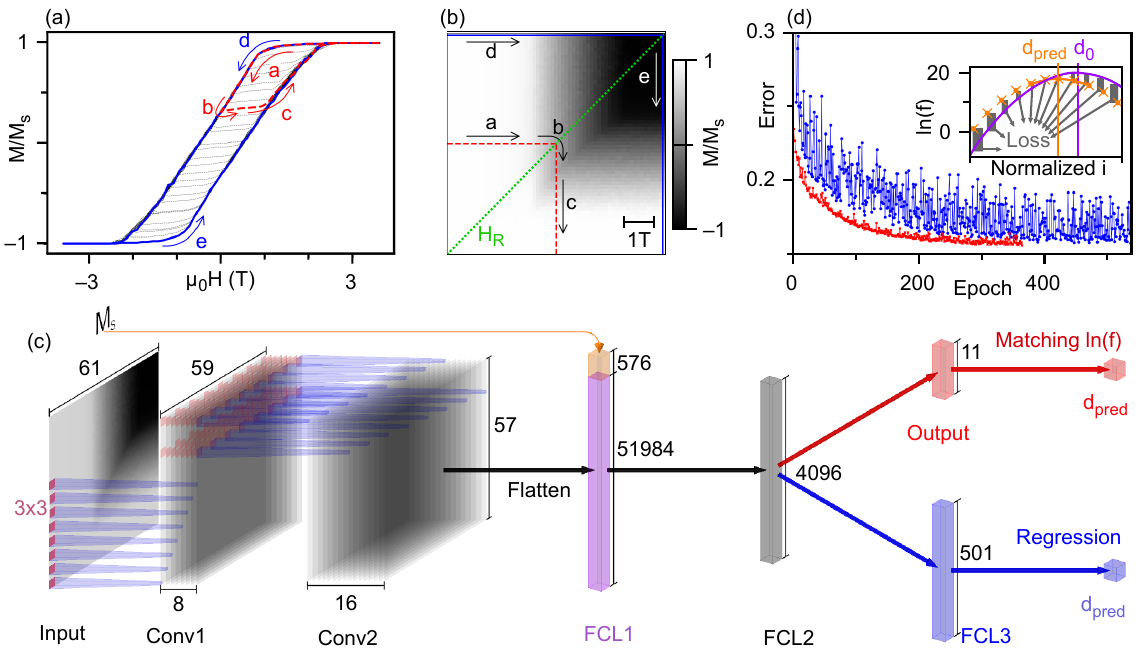}
\par\end{centering}
\caption{\textbf{Setup of the machine-learning problem.} (a) A typical family
of FORCs in our dataset. One minor loop is shown by the red dashed
line, whereas the full hysteresis loop is denoted by the solid blue
line. Other minor loops are illustrated by the gray dotted curves.
(b) The input image corresponding to the FORCs shown in (a), with
each pixel denoting one discrete point during the field scan. The
magnetization during the field scan is normalized to a 1-byte integer
between $0$ and $255$, corresponding to the brightness of each pixel.
(c) The structure of the CNN. The top path (red) denotes the distribution
matching method, whereas the lower path (blue) represents a conventional
regression. (d) The convergence of the training. The distribution-matching
and conventional regression results are denoted by the red and blue
curves, respectively. The inset illustrates the scheme of the loss
function in the distribution-matching method. \label{fig:FigureOneAboutFORCsAndCNN}}
\end{figure*}

In contrast to magnetic imaging, magnetometry is easily accessible,
e.g., via SQUID (superconducting quantum interference device), vibrating
sample magnetometry (VSM), or magneto-optical Kerr effect. However,
as conventional magnetometry typically measures ensemble-averaged
magnetic responses of a given system, it is not obvious that information
about the DMI can be extracted directly from magnetometry alone. Nevertheless,
magnetometry has been demonstrated to contain detailed information
about the magnetization reversal process. Specifically, the first-order
reversal curve (FORC) method, which utilizes many partial hysteresis
curves, has been used to ``fingerprint'' the magnetization reversal
process, providing information well beyond the ensemble average extracted
from conventional major hysteresis loop measurements \citep{pike_characterizing_1999,davies_magnetization_2004,gilbert_quantitative_2014}.

Here, we show that a convolutional neural network (CNN) can indeed
extract the DMI magnitude from FORCs. With a dataset of FORCs generated
for thin films via micromagnetic simulations, we demonstrate that
good accuracy can be achieved when the FORCs exhibit rich features.
The network performs particularly well when the SO(3) symmetry is
not significantly broken, i.e., when the uniaxial anisotropy is comparable
to the shape anisotropy of a thin film. This neural network provides
a convenient tool to extract DMI and investigate topological spin
textures in magnetic multilayers using magnetometry.

The FORC method measures a collection of many partial hysteresis curves,
which are known to contain important information about the distribution
of magnetic properties as well as interactions between magnetic elements
within a system. Typically, a sample is first saturated in a positive
field, then brought to a particular reversal field $H_{R}$, and the
magnetization $M$ is measured with increasing applied field $H$
back to positive saturation, thus tracing out a single FORC. The measurement
is repeated at successively more negative $H_{R}$, leading to a family
of FORCs. A second-order derivative is taken to extract a FORC distribution,
$\rho(H,H_{R})=-
(
\partial^{2}M(H,H_{R})/\partial{H}\partial{H_{R}})/2$ \citep{roberts_first-order_2000,davies_magnetization_2004,dobrota_what_2013}.
The FORC distribution has been used to probe irreversible magnetization
switching\citep{davies_magnetization_2004}, capture magnetic distributions\citep{rahman_controlling_2009,toro_remanence_2017,dumas_magnetic_2007},
distinguish different magnetic phases\citep{gilbert_realization_2015}
and switching mechanisms\citep{dumas_magnetic_2007}, and study magnetic
interactions\citep{gilbert_quantitative_2014,burks_3d_2020,biasi_quantitative_2016}.

While the FORC distribution mentioned above can be used to extract
useful information, the vast parameter space of many competing terms
in the Hamiltonian makes it difficult to interpret the DMI magnitude.
Specifically, the analysis that can extract the DMI magnitude from
FORCs is essentially unknown. However, this task can be easily converted
to a pattern recognition problem for artificial intelligence (AI).
With supervised learning, the first several convolutional layers in
a CNN can usually identify complicated, non-linear filters to extract
needed features autonomously with relatively low computational and
memory costs. To train such a CNN, we generated labeled data using
micromagnetic simulations.

Consider a phenomenological Hamiltonian
\begin{align}
H & =-J_{0}\sum_{\langle i,j\rangle}\vec{s}_{i}\cdot\vec{s}_{j}-\sum_{\langle i,j\rangle}D_{0}(\hat{z}\times\hat{r}_{ij})\cdot\left(\vec{s}_{i}\times\vec{s}_{j}\right),\nonumber \\
 & -K_{0}\sum_{i}(\vec{s}_{i}\cdot\hat{n})^{2}-\mu\sum_{i}(\mu_{0}\vec{H}_{\textrm{ext}}+\mu_{0}\vec{H}_{\textrm{dipole},i})\cdot\vec{s}_{i}\label{eq:FullHamiltonian}
\end{align}
where $\{\vec{s_{i}}\}$ denotes the normalized classical spins defined
on a square lattice $\{i\}$ in the $x$-$y$ plane, $\vec{r}_{ij}$
is the position vector connecting sites $i$ and $j$, $\hat{n}$
is the direction of the uniaxial anisotropy, and $\langle i,j\rangle$
denotes pairs of neighboring sites. The evolution of the spin texture
$\{\vec{s_{i}}\}$ governed by the Hamiltonian given by Eq.\ref{eq:FullHamiltonian}
can be simulated using micromagnetic simulations implemented by mumax$^{3}$\citep{vansteenkiste_design_2014},
taking $A_{ex}=J_{0}/2a$ as the exchange stiffness, where $J_{0}$
is the symmetric Heisenberg exchange energy and $a$ is the lattice
constant. Similarly $D=D_{0}/a^{2}$ is the DMI magnitude and $K_{u}=K_{0}/a^{2}t$
is the uniaxial anisotropy energy density in a film of thickness $t$.
Both the external magnetic field, $\vec{H}_{\textrm{ext}}$, and the
dipolar fields $\{\vec{H}_{\textrm{dipole},i}\}$ are included, assuming
periodic boundary conditions within the thin-film plane and a uniform
magnetic moment of $\mu=a^{2}tM_{s}$ within each site, where $M_{s}$
is the saturation magnetization. FORCs are then simulated by the average
spin of the ground state during the scan of $\vec{H}_{\textrm{ext}}$
applied perpendicular to the thin film. The thermal fluctuation of
finite temperature $T$ is included by a stochastic effective field. 

Each family of FORCs was simulated using a simulation box with an extent of $120\thinspace\textrm{nm}$ in the $x$ and $y$ directions split into a $16\times16$ grid with a periodic boundary condition of $5$ repeats in the plane of the film.
All samples had this geometry and used a damping factor of $\alpha=0.5$ to enable fast access to the ground state.
%To simulate disorder, both the magnitude of $K_{u}$ and the direction of the easy axis ($\hat{n}$) are varied within each simulation.
To simulate experimental uncertainty, both the magnitude of $K_{u}$ and the direction of the easy axis ($\hat{n}$) were given an extra layer of randomness.
The direction of the easy axis $\hat{n}(\theta,\phi)$ for the entire sample was chosen with the azimuthal angle $\phi$ being uniformly random within $[0,2\pi]$ and the polar angle $\theta$ chosen from a normal distribution centered at $0$ with a standard deviation of $\sigma_{\theta}$, which was itself chosen from a uniform distribution within $[0^{\circ}, 10^{\circ}]$.
The magnitude of $K_{u}$ for the sample was chosen according to a normal distribution centered at $K$ with a standard deviation of $\sigma_{K}$, which were themselves chosen from uniform distributions.
We used the Heun's method implemented in $\textrm{mumax}^{3}$ to solve the Landau-Lifshitz-Gilbert equation using adaptive time-stepping with $\textrm{MinDt}=2.5\times10^{-13}\thinspace\textrm{s}\,$ and $\textrm{MaxDt}=5\times10^{-12}\thinspace\textrm{s}$.
At every value of $\vec{H}_{\textrm{ext}}$, time was first progressed for $0.1\thinspace\textrm{ns}$ and then evolved in increments of $0.125\thinspace\textrm{ns}$ until $\langle s_{z}\rangle$ changed by less than $5\times10^{-3}\,$ between step increments. 
\begin{table}
\caption{\textbf{The bounds of simulation parameters}\label{table:theBounds}}

\centering{}%
\begin{tabular}{lcccccccc}
\toprule 
 & $M_{s}$ & $T$ & $A_{\textrm{ex}}$ & $K$ & $D$ & $t$ & $\sigma_{\theta}$ & $\sigma_{K}$\tabularnewline
\, & A/m & K & J/m & J/m$^{3}$ & J/m$^{2}$ & nm & $^{\circ}$ & $\%$\tabularnewline
\midrule 
Min & $\,\,\,\,2\times10^{5}$ & $\,\,\,\,\,\,135$ & $\,\,\,1\times10^{-12}$ & 0 & 0 & $\,\,10$ & 0 & 0\tabularnewline
Max & $14\times10^{5}$ & 19420 & $35\times10^{-12}$ & $10^{6}$ & 0.005 & 100 & 10 & 20\tabularnewline
\bottomrule
\end{tabular}
\end{table}

\begin{table}
\centering{}\caption{\textbf{Parameters of our CNN.} \label{tab:CNN_Details}}
\begin{tabular}{cp{1.7cm}p{1.7cm}cr}
\toprule 
Layer & \centering{}Input & \centering{}Output & Kernel Size & Dropout\tabularnewline
\midrule 
Conv1 & \centering{}$61\times61$ & \centering{}$59\times59$ & $3\times3$ & 0\tabularnewline
\midrule 
Conv2 & \centering{}$59\times59$ & \centering{}$57\times57$ & $3\times3$ & 0.5\tabularnewline
\midrule 
FCL1 & \centering{}$52,560$ & \centering{}$4,096$ & \multicolumn{1}{c}{N/A} & 0.5\tabularnewline
\midrule 
FCL2 & \centering{}$4,096$ & \centering{}$11$ & \multicolumn{1}{c}{N/A} & 0.5\tabularnewline
\bottomrule
\end{tabular}
\end{table}

To interface FORCs with a CNN, we convert each family of FORCs {[}one
typical example shown in Fig. \ref{fig:FigureOneAboutFORCsAndCNN}(a){]}
to an information-dense input image {[}Fig. \ref{fig:FigureOneAboutFORCsAndCNN}(b){]}.
First, the full range of scanning magnetic field $[H_{\textrm{min}},H_{\textrm{max}}]$
is discretized to $61$ steps. A family of FORCs can then be rearranged
as a $61\times61$ image, with each pixel denoting one discrete step
of the field scan. During the scan, the normalized magnetization value
given by $\langle s_{z}\rangle$ is discretized to integers within
$[0,255]$, which are denoted by the brightness of the pixels in Fig.
\ref{fig:FigureOneAboutFORCsAndCNN}(b). For each minor loop in Fig.
\ref{fig:FigureOneAboutFORCsAndCNN}(a), the applied field starts
from positive saturation $H_{\textrm{max}}$ (the top right corner),
then hits the reversal field $H_{R}$ and eventually goes back to
positive saturation (red arrows $a\rightarrow b\rightarrow c$). Correspondingly,
the pixel in Fig. \ref{fig:FigureOneAboutFORCsAndCNN}(b) first scans
horizontally from the left to the right (arrow a). When the scanning
field in Fig. \ref{fig:FigureOneAboutFORCsAndCNN}(a) reaches $H_{R}$
the scanning pixel hits the diagonal green dotted line in Fig. \ref{fig:FigureOneAboutFORCsAndCNN}(b).
This pixel then scans vertically downwards (arrows $b\rightarrow c$)
as the field in Fig. \ref{fig:FigureOneAboutFORCsAndCNN}(a) scans
back to positive saturation. With this arrangement, each minor loop
traced out by a FORC in Fig. \ref{fig:FigureOneAboutFORCsAndCNN}(a)
is mapped to a horizontal and a vertical segment of pixels in Fig.
\ref{fig:FigureOneAboutFORCsAndCNN}(b) connected at the dotted green
line. When $H_{R}=H_{\textrm{max}}$, the minor loop in Fig. \ref{fig:FigureOneAboutFORCsAndCNN}(a)
shrinks to zero, contributing only a bright pixel at the lower left
corner in Fig. \ref{fig:FigureOneAboutFORCsAndCNN}(b). As $H_{R}$
becomes more negative, the minor loops in Fig. \ref{fig:FigureOneAboutFORCsAndCNN}(a)
enlarge, and eventually recover the full hysteresis loop when $H_{R}=H_{\textrm{min}}$.
Correspondingly, the path in Fig. \ref{fig:FigureOneAboutFORCsAndCNN}(b)
gradually becomes larger and eventually completes the image by filling
the outermost row and column of pixels (d$\rightarrow$e). In our
simulations, the range of the scan is $[-3.6\thinspace\textrm{T},+3.6\thinspace\textrm{T}]$.
Each pixel in the image thus corresponds to one step of $0.12\thinspace\textrm{T}$.
This choice of step size balances feature resolution and simulation
time.

To ensure data diversity, each family of simulated FORCs is parameterized
by $\{M_{s}$, $T$, $A_{\textrm{ex}}$, $K_{u}$, $D$, $t$, $\theta$, $\phi$ $\}$, with each element being a double-precision floating-point number randomly generated
using the values and bounds shown in Table \ref{table:theBounds}. The
normalized DMI magnitude $d=\frac{D-D_{\textrm{min}}}{D_{\textrm{max}}-D_{\textrm{min}}}$
is used as the label. We determined the range of these material parameters
based on reported experimental values. For $M_{s}$, we set the upper
bound according to bulk cobalt, whereas the lower bound is set to
mimic the reduced value in typical magnetic multilayers with non-magnetic
or antiferromagnetic components \citep{shaw_reversal_2008,boulle_room-temperature_2016,gilbert_realization_2015,shim_ultrafast_2020,heigl_dipolar-stabilized_2021}.
$A_{\textrm{ex}}$ is set to vary between the cobalt value obtained
in first-principles calculations \citep{moreno_temperature-dependent_2016}
and those values previously obtained in soft magnetic multilayers
or alloys \citep{eyrich_exchange_2012,pollard_observation_2017}. The
value of $K$ is set between zero and a typical value for a Co/Pd multilayer
with strong perpendicular magnetic anisotropy \citep{shaw_reversal_2008,gilbert_realization_2015}.
As the main target of this work we set the DMI magnitude from zero
to $5\times10^{-3}\thinspace\textrm{J/m}^{2}$, an upper bound exceeding
most observed DMIs in multilayers \citep{boulle_room-temperature_2016,moreau-luchaire_additive_2016,soumyanarayanan_tunable_2017}.
We chose the bounds of temperature based on a rough estimate of the
Curie temperature: $T_{C}\approx\frac{A_{\textrm{ex}}a}{k_{B}}$,
such that $T\in[\min(T_{C})-\Delta T,\max(T_{C})+\Delta T]$, leaving
margins of $\Delta T=\frac{3}{4}\min(T_{C})$ to sample the cases
of $T<T_{C}$ and $T>T_{C}\,$. Note that these 
are the simulation temperatures. In micromagnetic simulations,
the coarse-graining process leads to an effective physical temperature that is rescaled from the simulation temperature. 
In our case, the maximum effective temperature is estimated to be
roughly 10 to 20 times smaller than the maximum simulation
temperature \citep{grinstein_coarse_2003, hahn_temperature_2019}.

We utilized Pytorch, a machine learning framework
with a Python interface, to construct and train our CNN.
The specific model of our CNN is illustrated in Fig. \ref{fig:FigureOneAboutFORCsAndCNN}(c),
where two convolutional layers ($\textrm{Conv1}$ and $\textrm{Conv2}$)
and two fully connected layers ($\textrm{FCL1}$ and $\textrm{FCL2}$)
are employed with ReLU (Rectified Linear Unit) activation function.
The details of the model are listed in Table \ref{tab:CNN_Details}.
During training, a dropout rate of $0.5$ was applied to $\textrm{Conv2}$
and all fully connected layers. 
The configuration
of our CNN was informed by a series of trial-and-error tests in which we varied, among other things, 
%the product of trial and error to nd which setup gave the best results. Among other things, we varied 
the number of convolutional layers, the amount
of feature maps in each layer, the size and number of fully
connected layers, the activation functions between the
layers, and the combination of dropout rates used for the
layers. Similarly, we surveyed the parameters of the ADADELTA adaptive learning rate method, and eventually used a momentum of $\rho=0.9$, a learning rate of $\gamma=0.01$, and a weight decay
of $\lambda=0$ \citep{zeiler_adadelta_2012}. 
%in much the same way as the CNN architecture.}

In practice, since $M_{s}$ is readily available
from typical magnetometry measurements, it is feasible to include
it, along with the normalized FORCs, as input to the neural network.
This is similar to the idea of residual network \citep{He_2016} where
additional information can skip some sections of the network to mimic
long-term memory. In our CNN, $M_{s}$ is first normalized to $[0,10]$
and then directly fed to $\textrm{FCL1}$ by appending $576$ nodes
of the same value, along with the flattened feature maps extracted
by all the convolutional layers. The number of $M_s$ nodes was large enough to avoid concerns of being dropped out while being small to not overwhelm the feature map information.

Different from a typical regression
model, we implemented a distribution matching method as described
by Eqs. \ref{eq:y}-\ref{eq:prediction}, 
\begin{eqnarray}
\vec{y} & = & \exp{\left[\vec{{\cal {F}}}_{\textrm{CNN}}(\vec{x})\right]},\label{eq:y}\\
{\cal {L}} & = & \left\langle \left[\ln\vec{y}-\ln\vec{f}(d_{0})\right]^{2}\right\rangle ,\label{eq:lossFunction}\\
d_{\textrm{pred}} & = & \underset{d}{\arg\min}'\left[\ln\vec{y}-\ln\vec{f}(d)\right]^{2}\label{eq:prediction}
\end{eqnarray}
where $\vec{{\cal {F}}}_{\textrm{CNN}}(\vec{x})$ is the CNN output
given input $\vec{x}$. Unlike conventional regression method, the
output involves more nodes than needed (in our case $11$). We then
map the output to positive values $\vec{y}>0$, and train the CNN
by matching $\vec{y}$ to some smooth distribution function $\vec{f}(d)>0$
centered at $d$ by using the loss function ${\cal {L}}$ defined
in Eq. \ref{eq:lossFunction}, where $d_{0}$ is the label. When making
the prediction, the value of $d_{\textrm{pred}}$ is determined by
an argument minimization searching for the best fit between $\vec{f}(d)$
and $\vec{y}$. The prime in Eq.\ref{eq:prediction} denotes that
the search for $d_{\textrm{pred}}$ involves only certain ($n$) output
nodes centered at $\underset{i}{\arg\max}(\vec{y})$. This distribution
matching method allows not only for a bounded, continuous prediction
from a discrete output, but it also enables the possibility to find
the confidence of the prediction $\xi[\vec{y},d_{\textrm{pred}}(\vec{y})]$
using the redundant output information, as will be discussed later.

In practice we used Gaussian $f_{i}=e^{C}e^{-\frac{1}{2}(\frac{i-d}{\sigma})^{2}}$
for $\vec{f}(d)$, where $i$ is the index of the output layer normalized
within $[0,1]$, and $(C=20,\sigma=0.1,n=5)$ are chosen for the coefficients.
This makes Eq. \ref{eq:prediction} equivalent to the search for a
best-fit parabola based on five given points {[}Fig. \ref{fig:FigureOneAboutFORCsAndCNN}(d)
inset{]}, for which the formalism is straightforward (See Supplementary
Information). In principle one could train a deep neural network to
replace Eq. \ref{eq:prediction} and the model would become a standard
regression \cite{Khoo_quantum_2021}. To show this, we trained an alternative model by first
feeding FCL2 to another fully connected layer, FCL3, with $501$ nodes,
from which the prediction is directly extracted, as illustrated by
the lower path in Fig. \ref{fig:FigureOneAboutFORCsAndCNN}(c). The
convergence paths of the two models are compared in Fig. \ref{fig:FigureOneAboutFORCsAndCNN}(d),
where the standard regression model converges more slowly and exhibits
more significant fluctuation of the error.

\begin{figure}
\begin{centering}
\includegraphics[width=1\columnwidth]{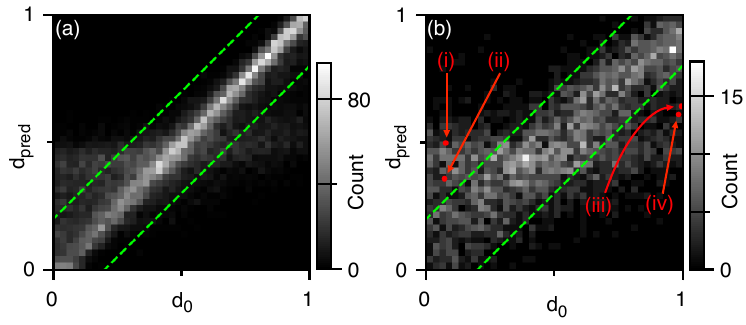}
\par\end{centering}
\caption{\textbf{Performance of the CNN.} (a-b) The statistics of the training
outcome among the training set (a) and the testing set (b). The dashed green lines denote a prediction error of 20\%
marking our threshold for good guesses. The red arrows
in (b) denote four representative cases (i-iv) among the faint band
of wrong predictions. \label{fig:TrainningAndTestingResults}}
\end{figure}

To train the model, we generated $20,000$ families of FORCs with
recorded simulation parameters. Several examples of FORCs are shown
in Supplementary Information. For the training we used $80\%$ of
the data separated into $32$ batches, which were shuffled after all
batches are accessed during each training epoch. The rest of the data
was held out as a test set, which was used to examine the performance
after each epoch. The training continued until the best performance
was not superseded for 200 epochs straight. After training, the average 
absolute error was 
$\sim0.096$ and $\sim0.155$ for the training and
the testing sets, respectively. The statistics for these outcomes
are shown in Figs. \ref{fig:TrainningAndTestingResults}(a-b), where
the bright major diagonal lines in both results correspond to accurate
predictions. This indicates that the CNN can indeed extract the magnitude
of DMI from FORCs well beyond random guessing. Further training can
indeed improve the performance for the training set, but not for the
test set, consistent with the trend expected for overfitting. Besides
the main diagonal lines, a faint, nearly flat background band also
shows up in both results, suggesting that the CNN is unable to make
the correct predictions in certain cases. 
\begin{figure}
\begin{centering}
\includegraphics[width=1\columnwidth]{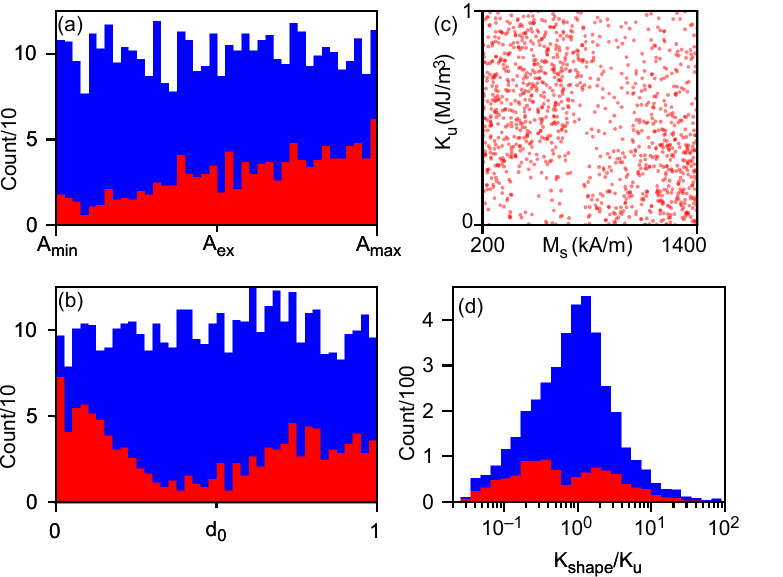}
\par\end{centering}
\caption{\textbf{The statistics of the predictions.} The distributions of $A_{\textrm{ex}}$
and $d_{0}$ among the testing data are shown in (a) and (b), respectively.
The red histogram denotes the predictions with errors greater than
$20\%$, whereas the blue ones illustrate the entire test set. (c)
The distribution of $M_{s}$ and $K_{u}$ among the wrong predictions.
(d) The distribution of $\frac{K_{\textrm{shape}}}{K_{u}}$ among
the wrong predictions (red) and the entire test set (blue).\label{fig:StatisticsInErrorBands}}

\end{figure}

To understand the origin of the faint nearly-flat band in Figs. \ref{fig:TrainningAndTestingResults}(a-b),
we compare the simulation parameters for the group of FORCs with ``wrong" predictions to those
for the entire data set. Here, we define ``wrong" predictions to be those where
$|d_{\textrm{pred}}-d_{0}|\geqslant0.2$ {[}i.e. those points not
between the green dashed lines in Fig. \ref{fig:TrainningAndTestingResults}(b){]}. Since the parameters
are uniformly random, the test set is well balanced for all simulation
parameters. This can be seen from the total histograms (blue) of $A_{\textrm{ex}}$
and $d_{0}$ in Figs \ref{fig:StatisticsInErrorBands}(a) and (b),
respectively. However, among the wrong predictions (red) a skew towards
larger values of $A_{\textrm{ex}}$ can be clearly seen. This is intuitive
considering that the Heisenberg exchange is the symmetric part of
the exchange coupling that directly competes with the antisymmetric
DMI. On the other hand, the histogram of $d_{0}$ has a dip near the
center and rises near the edges, which is consistent with the faint
flat band of wrong predictions in Figs. \ref{fig:TrainningAndTestingResults}(a-b).
\begin{figure}
\begin{centering}
\includegraphics[width=1\columnwidth]{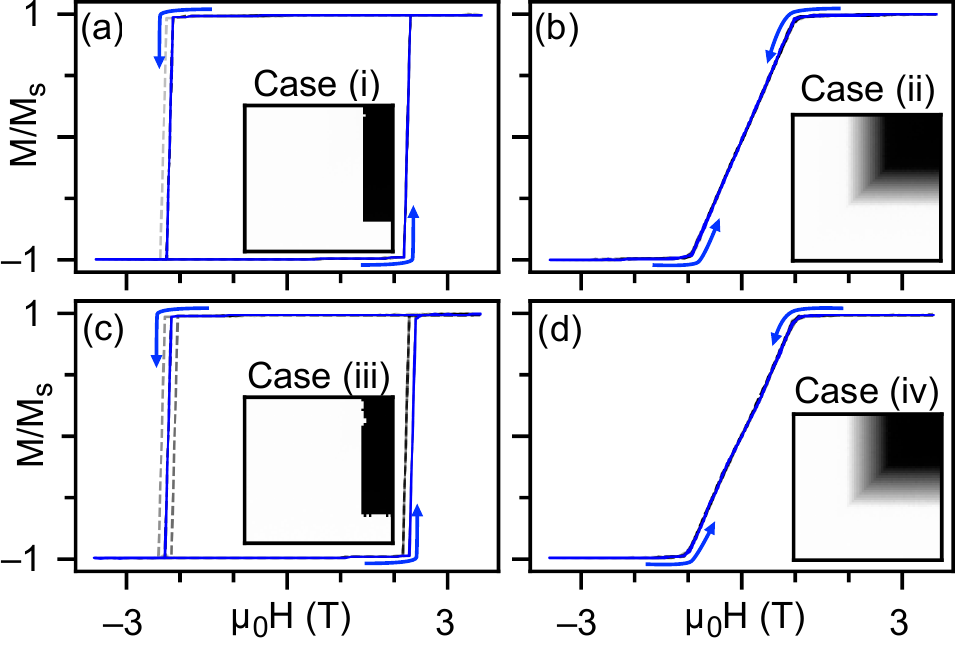}
\par\end{centering}
\caption{\textbf{Example inputs corresponding to uncertain predictions.} (a-d) The hysteresis loops for Cases
(i-iv) highlighted in Fig. \ref{fig:TrainningAndTestingResults}(b).
The inset in each case illustrates the corresponding input image fed
to the CNN.
\label{fig:FigureThreeFourCases}}
\end{figure}

The main reason of the prediction uncertainty can be uncovered by
comparing the statistics of $M_{s}$ and $K_{u}$ for the wrong predictions
among the test set. As one can see in Fig. \ref{fig:StatisticsInErrorBands}(c),
the wrong predictions are heavily populated where either $K_{u}$
is large and $M_{s}$ is small or vice versa. This suggests that the
competition between the easy-axis and the shape anisotropy plays an
essential role. When magnetized uniformly, the demagnetization field
contributed by the long-range dipolar interaction can be effectively
seen as an easy-plane shape anisotropy with $K_{\textrm{shape}}=\frac{1}{2}\mu_{0}M_{s}^{2}$.
This results in a perpendicular hard axis that competes with the easy
axis intrinsically hosted by the material. When these two terms almost
cancel in the Hamiltonian, for each site the SO(3) rotation symmetry
is restored in the case of uniform spin. The average magnetization
thus depends more on the exchange interactions between neighboring
sites. This makes the competition between $A_{\textrm{ex}}$ and $D$
more pronounced, which makes it easier for the CNN to recognize the
features of the DMI. Fig. \ref{fig:StatisticsInErrorBands}(d) shows
histograms of $K_{\textrm{shape}}/K_{u}$ for the full test set (blue)
and the erroneous portion (red), which demonstrates that the CNN has
more difficulty when either anisotropy dominates. This difficulty
originates from the fact that the FORCs are featureless in either
case. This can be seen in four representative FORCs {[}Figs. \ref{fig:FigureThreeFourCases}(a-d){]}
corresponding to the wrong predictions highlighted in Fig. \ref{fig:TrainningAndTestingResults}(b)
as Cases (i)-(iv). When $K_{u}$ is very large (Cases i and iii),
the FORCs switch abruptly such that almost no minor loops show up
within our 61-step resolution of $H_{R}$. On the other hand, when
$K_{\textrm{shape}}$ dominates (Cases ii and iv), the switching is
gradual but the hysteresis vanishes, leaving a featureless curve that
carries no information of the detailed spin texture. These FORCs remain
featureless for both large and small values of $D$, suggesting that
the DMI information is indeed overwhelmed. 

\begin{figure}
\begin{centering}
\includegraphics[width=1\columnwidth]{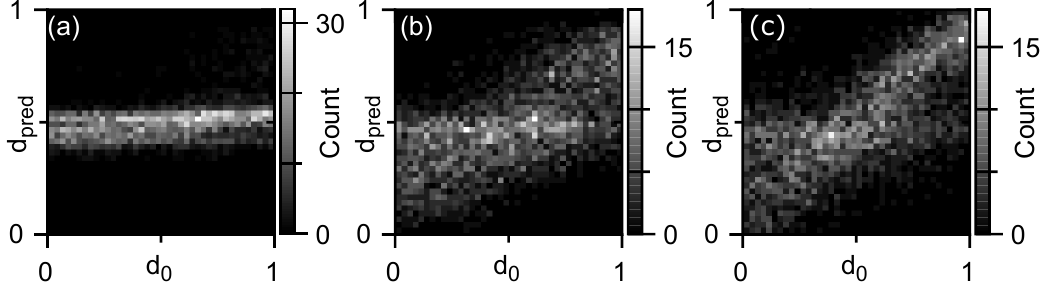}
\par\end{centering}
\caption{\textbf{Evolution of CNN performance.} (a-c) The evolution of the CNN performance during training.
The three panels (a-c) correspond to Epochs 3, 50 and 385, respectively.\label{fig:TrainingMontage}}
\end{figure}

The faint flat band formed by the wrong predictions in Figs. \ref{fig:TrainningAndTestingResults}(a)
and (b) can be understood by checking the learning strategy during
training. At different stages of training the statistics on the testing
set are recorded, among which three representative snapshots are shown
in Figs. \ref{fig:TrainingMontage}(a-c).  It seems that the CNN
first learns to guess around a single number because even this na\"ive
strategy reduces the average loss from random guesses by a factor
of $5$. After this, it starts to recognize the patterns associating
the FORCs with the DMI and gradually forgets the na\"ive strategy.  As
training continues, the CNN starts to make more accurate predictions
for nearly all FORCs with better pronounced features, leaving the
featureless FORCs in the faint band of wrong guesses due to the memory
of the na\"ive strategy.  Note that further training can indeed remove
this memory for the training data, whereas the performance on the
testing data does not improve, suggesting overfitting. A video showing
the evolution of the CNN performance can be found in Supplementary
Information.

\begin{figure}
\begin{centering}
\includegraphics[width=1\columnwidth]{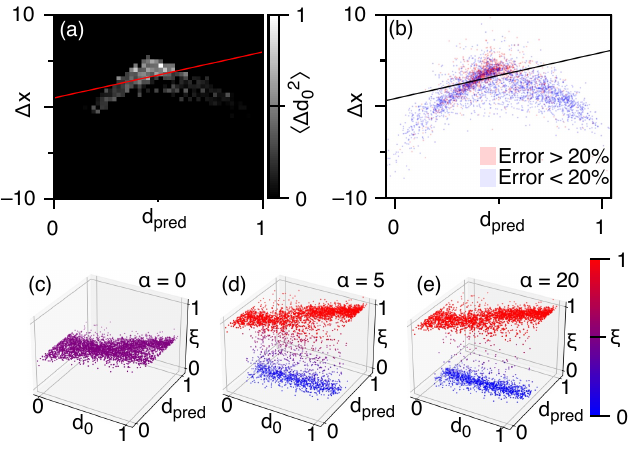}
\par\end{centering}
\caption{\textbf{Confidence metric established from the distribution-matching
method.} (a) The distribution of the variance of $d_{0}$ among the
test set illustrated as a function of $\Delta x$ and $d_{\textrm{pred}}$,
where the red solid line denotes our selected boundary separating
confident and uncertain predictions. The statistics within each pixel
are only shown if the sample size is more than $7$ points. (b) The
absolute error distribution of all predictions made among the test
set. The dark line in (b) is the same as the red solid line in (a).
(c-e) the separation between confident and uncertain predictions by
modulating the threshold coefficient $\alpha$. \label{fig:FigureFourConfidence}}
\end{figure}

Our CNN is different from a standard regression model since it interprets
the prediction using the distribution matching method mentioned above.
The redundant output information can be used to identify a function
$\xi[\vec{y},d_{\textrm{pred}}(\vec{y})]$ that extracts the confidence
of each prediction without knowing the correct answer $d_{0}$. Although
classification models naturally have this capability, the confidence
interpretation for a typical regression model is non-trivial \citep{barber_predictive_2021}.

In our particular case, we empirically observed that the average log
distance between the proper distribution and the output ($\Delta x=\langle\ln{[\vec{f}(d_{\textrm{pred}})]-\ln{\vec{y}}}\rangle$)
is strongly correlated with $d_{\textrm{pred}}$. This can be seen
in Fig. \ref{fig:FigureFourConfidence}(a), where the variance of
$d_{0}$ is illustrated for different combinations of $\Delta x$
and $d_{\textrm{pred}}$. We can then identify a linear function that
separates the confident and uncertain predictions, $\Delta x=\frac{d_{\textrm{pred}}}{0.2}+0.9$,
as shown by the red line in Fig. \ref{fig:FigureFourConfidence}(a).
To further examine the validity of such observations, we illustrate
the relation between $\Delta x$ and the absolute error in Fig. \ref{fig:FigureFourConfidence}(b),
where the confident predictions are mainly contributed by those with
errors less than $20\%$. Note that we do have correct predictions
above the dark line in Fig. \ref{fig:FigureFourConfidence}(b), which
are essentially lucky guesses since the corresponding variance is
large, as shown in Fig. \ref{fig:FigureFourConfidence}(a). Finally,
we define the overall metric of confidence within $(0,1)$ by $\xi[\vec{y},d_{\textrm{pred}}(\vec{y})]=\sigma[-\alpha(\Delta x-\frac{d_{\textrm{pred}}}{0.2}-0.9)]$,
which is a sigmoid function centered at the red line in Fig. \ref{fig:FigureFourConfidence}(a),
with the tolerance modulated by $\alpha$. As shown in Figs. \ref{fig:FigureFourConfidence}(c-e),
when modulating $\alpha$ one can eventually separate the confident
predictions from the entire test set. This is not possible with a
conventional regression model where the only output information is
the prediction. Note that $\xi[\vec{y},d_{\textrm{pred}}(\vec{y})]$
is not unique and the boundary of confident predictions is not necessarily
a linear function. In principle, one can build another neural network
to find a $\xi[\vec{y},d_{\textrm{pred}}(\vec{y})]$ that performs
better, which is beyond the scope of this work.

To conclude, we demonstrate that the magnitude of the DMI is indeed
contained in the hysteretic magnetometry data. Without any information
of spin-texture details, a CNN can recognize the DMI magnitude from
feature-rich FORCs with an error of $\sim15\%$. The prediction is
particularly confident when the intrinsic perpendicular easy-axis
anisotropy is comparable to the shape anisotropy, which can be fine-tuned
in experiments. This brings about the possibility to evaluate the
magnitude of DMI without any spin-texture characterization. 
Future directions to improve accuracy in applications to experimental data include the generation of higher resolution simulation data within parameter ranges of particular experimental interest, as well as the injection of noise into the training data.  Our results also suggest that there may exist a model that relates the DMI magnitude
with the ensemble-averaged magnetization, which invites further theoretical
investigations.

The database ``HoyaFORCs'' and the example code to train the CNN
are available at https://github.com/bfugetta/HoyaFORCs
\begin{acknowledgments}
\emph{Acknowledgments:} This work has been supported by the US National
Science Foundation grants DMR-1950502 (BF, AYL), DMR-2005108 (ZC,
DB, KL), and ECCS-2151809 (GY, KL).
\end{acknowledgments}

\end{document}